%% file: main.tex
\documentclass[letter,12pt]{article}

\usepackage[top=1in,left=1in,right=1in,bottom=1in]{geometry}

\input{preamble}

\usepackage[affil-it]{authblk}

\newcommand{\keywords}[1]{\par\addvspace\baselineskip\noindent\textbf{Keywords:}\,#1}

\title{\revisitingTitle}
\author[1]{Samuel J. Eschker}
\author[1]{Chuanhai Liu \footnote{Corresponding author: \revisitingCorrespondenceEmail}}

\affil[1]{\it Department of Statistics, Purdue University, West Lafayette, IN 47907, USA}

\date{June 19, 2026}

\begin{document}

\input{revisiting_stein_paradox_body}

\end{document}

%% file: revisiting_stein_paradox_body.tex
\maketitle

\begin{abstract}
  The classical multivariate normal means problem remains conceptually
  unresolved. While shrinkage and empirical Bayes methods
  improve risk by imposing external geometric or hierarchical structure,
  they fail to explain how information is shared
  across independent coordinates for a fixed, unstructured mean vector. We
  address this gap using the prior-free Inferential Models framework.
  By formulating a generalized probability integral transform (GPIT)
  for independent, non-i.i.d~observations combined with a
  reweighted Anderson-Darling predictive random set, we leverage the global 
  shape of ordered observations for valid, efficient inference. Crucially, the 
  auxiliary structure of this formulation provides a novel explanation for 
  Stein's paradox, demonstrating that the maximum likelihood estimator 
  becomes structurally implausible for $n\geq 3$. To ensure scalability, we 
  introduce an i.i.d.~sampling-with-replacement surrogate that connects our 
  exact fixed-mean formulation to overparameterized $g$-modeling. Furthermore, we develop a maximin criterion for combining plausibility contours.
  Under squared error loss, our estimators are competitive
  with state-of-the-art auto-modeling methods and 
  outperform classical shrinkage and empirical Bayes methods.   
\end{abstract}


\section{Introduction}
\label{s:intro}

The multivariate normal means problem concerns simultaneous
estimation of
$\theta\in\mathbb R^n$ from a single observation
$
  X\sim N_n(\theta,I_n)
$
where $\theta$ is a fixed, unknown parameter or, in the
empirical-Bayes variant called $g$-modeling,
$\theta_i \stackrel{\text{i.i.d.}}{\sim} G$ for some unknown distribution $G$.
The maximum likelihood estimator (MLE), $\hat\theta=X$, would seem to be a natural choice because the $X_i$ are independent and marginally, $X_i$ is complete, sufficient, the UMVUE, and minimax for $\theta_i$. Nevertheless, \citet{Stein1956} established that the MLE is
inadmissible under squared error loss for $n\geq 3$. The James--Stein estimator
\citep{js1961}, empirical Bayes methods \citep{efron1972empirical}, and
modern shrinkage procedures all leverage the joint structure of the
coordinates to reduce risk. However, each of these methods
imposes additional structure in the estimation procedure that deviates from
the classical, fixed-mean problem, leaving the original problem conceptually unresolved \citep{stigler1988NeymanMemorial1990}.

In this paper, we propose a class of point estimators built upon the
prior-free foundation of Inferential Models (IMs)
\citep{martin2013inferential, martin2015inferential}, which
\citet{cuiDemystifyingInferentialModels2025a}
identify as ``one of the
original statistical innovations of the 2010s.'' IMs represent
the sampling model as an association
that relates the data, the parameter, and an auxiliary variable with a known
distribution. Our central claim is that reasoning with a carefully chosen
auxiliary distribution captures the underlying global structure
of the high-dimensional parameter $\theta$ (as revealed by the ordered sample)
without imposing additional structure on the model. Consequently, by analyzing
the radial auxiliary structure (Section \ref{s:radial-scores}), this framework provides a novel structural explanation of the failure of the MLE when
$n\geq 3$, shining new light on Stein's paradox.

To build such an association, we develop a generalized probability integral
transform (GPIT) which maps the order statistics of the independent,
non-i.i.d.~sample under the
true parameter to the order statistics of an i.i.d. uniform sample. This
generalizes the classical probability integral transform and
provides a rank-preserving, bijective mapping
to an ordered-uniform distribution under the true $\theta$. This
GPIT allows us to utilize, as our auxiliary variable,
a one-sample test statistic based on a reweighted
Anderson--Darling statistic \citep{liu2023reweighted}, which is specifically
tailored for tail-sensitive departures from the
ordered-uniform law. Paired with an valid predictive random set,
this association captures the global shape structure of $\theta$,
enabling the validation or rejection of candidate parameter vectors based on 
how well the reconstructed auxiliary variables conform to the ordered uniform 
law.

The outcome of this initial prediction is a set of $\theta$, whereas the
estimation problem requires all $\theta_i$ to be jointly associated with
$X_i$ by predicting an associative permutation.
Evaluations of the exact distribution to
resolve this assignment require summing over all $n!$ latent
permutations of parameters to coordinates. While this exact evaluation is strictly computable for small dimensions (and we demonstrate its efficacy for $n\leq 8$), it is infeasible for large $n$.
We resolve this scaling limitation for a fixed $\theta$ by introducing
a second setup: a large-$n$ sampling-with-replacement
surrogate. This surrogate replaces sampling-without-replacement from the
coordinates of $\theta$ with an independent and identically
distributed model corresponding to sampling with
replacement from an empirical mixture distribution. By transitioning to this
i.i.d.~model, we bypass the need for expensive assignment sums with
only a small increased-risk cost. In simulation studies,
this approximation procedure performs as well as state-of-the-art
methods with increasing $n$.

Our third setup extends this i.i.d. approximation from the fixed $\theta$ construction to an overparameterized empirical Bayes variant, known as
$g$-modeling \citep{robbinsEmpiricalBayesApproach1956,efron_two_2014},
\begin{equation}\label{eq:g-modeling}
  X_i \stackrel{i.i.d.}{\sim} \int_{-\infty}^\infty \phi(x-\theta)d
  G(\theta),\qquad x\in \mathbb{R},\; i=1, ..., n,
\end{equation}
where the unknown distribution $G$ is represented flexibly.
This may be a scientifically meaningful modification of the model
for certain applications, and this approximation preserves the first-order
empirical mixture generated by $\theta_1,\dots,\theta_n$ while modeling the entire distribution $G$.
Classically, the log-density of $G$ has been estimated by applying
penalized maximum likelihood to the coefficients of a low-dimensional
spline basis \citep{efron2016empirical}, or G is estimated directly
through the non-parametric MLE
\citep{lairdNonparametricMaximumLikelihood1978,jiang2009gmleb}.
\citet{efronBayesOracleBayes2019} and \citet{zhang2003compound}
review the developments of the compound decision and oracle Bayes methods.
Instead, we estimate $G$ using over-parameterized IMs, furthering recent developments in over-parameterized $g$-modeling \citep{jiang2025estimation}.

Through these developments, our primary contributions are theoretical, methodological, and computational. The theoretical contributions are twofold.
First, we construct a class of point
estimators that capture global structure in high dimensions through
a rank-preserving, bijective GPIT. We show that this construction allows
for valid and efficient inference.
Second, we provide a novel structural
explanation of why the MLE fails when $n \geq 3$, complementing
decision-theoretic, Bayesian, and
geometric explanations of Stein's paradox. Specifically, we show how
the MLE corresponds to an
atypical, zero-density point of the joint auxiliary variable
distribution, establishing
its implausibility from an auxiliary perspective. Methodologically, we 
introduce a procedure for combining multiple distinct plausibility
contours using a maximin (bottleneck) criterion and devise a method for 
fitting overparameterized $g$-modeling using IM. Computationally, 
we establish a sampling-with-replacement approximation to the exact GPIT 
construction for large $n$. Extensive numerical experiments 
verify that our new estimators, arising from both the exact and approximate
constructions, perform competitively with state-of-the-art
auto-modeling estimators and outperform James--Stein and empirical Bayes
methods across a wide range of structural settings.

The remainder of this paper is organized as follows.
Section~\ref{s:IMs} reviews IMs and introduces the
ordered-uniform predictive random
set based on the reweighted Anderson--Darling statistic.
Section~\ref{s:cmnm} develops the exact GPIT
construction for the classical high-dimensional normal mean problem,
details point estimation, and
presents the large-$n$ sampling surrogate. Section~\ref{s:composite}
proposes a procedure to combine multiple plausibility contours.
Section~\ref{s:eb}
develops the corresponding IM
construction for the over-parameterized $g$-modeling case.
Section~\ref{s:numerical} presents comprehensive
numerical comparisons, and Section~\ref{s:discussion} concludes with a
discussion.

\section{Inferential models and prediction of ordered uniforms}
\label{s:IMs}

\subsection{Review of Inferential Models}\label{s:ims}

An IM is a framework for prior-free probabilistic inference
motivated by fiducial and Dempster-Shafer methods.
It associates unknown parameter values
$\theta\in\Theta$ to observed data $X \in {\mathbb X}$ through auxiliary
variables $U$ with known distribution
${\mathbf P}_U$ via an {\it association}, $a(X, U, \theta) = 0$ where 
  $X \in {\mathbb X}$, $\theta\in\Theta$, and $U \sim {\mathbf P}_U$.
To perform inference, we utilize a valid (frequency-calibrated)
predictive random set $\mathcal{S}$ for $U$.
Formally, a predictive random set $\mathcal{S}$ is {\it
valid} if it is independent of $U$ and
\begin{equation}\label{eq:valid-prs}
  \mbox{Prob}[\mbox{Prob}(\mathcal{S} \not\ni u) \ge 1-\alpha]
  \le \alpha \quad \forall \alpha \in (0,1).
\end{equation}
If the outer inequality can be placed
with equality, then we say $\mathcal{S}$ is {\it efficient}.
Typically, $\mathcal{S}$ is chosen in such a way that
$\mbox{Prob}(\mathcal{S} \not\ni u) \sim \mbox{Uniform}(0,1)$
(c.f.~conditional and marginal IMs of
  \citet{martin2015conditional, martin2015marginal},
  but sufficient in the present case of simultaneous inference about
$n$ parameters $\theta_i$ from $n$ observations with $n$ auxiliaries).
Through the inverse set-valued mapping given by the association, this
predictive random set induces the random set on $\Theta$, ${\Theta_X(\mathcal{S}) =\{\theta: a(X, u, \theta)=0 \,\mbox{ for some }\,
  u \in \mathcal{S}\}}$.

Throughout, we assume $\Theta_X(\mathcal{S})\neq \emptyset$
for all values of $\mathcal{S}$
(cf.~\citet{leaf2012inference} for the case
where $\mbox{Prob}(\mathcal{S} = \emptyset)>0$).
Uncertainty regarding an assertion $A \subseteq \Theta$, is quantified by belief, $$\mbox{bel}_X(A) = \mbox{Prob}(
    \Theta_X(\mathcal{S}) \subseteq A
  ),$$ and plausibility, $${\mbox{pl}_X(A) = 1 - \mbox{bel}_X(A^c) = \mbox{Prob}(
    \Theta_X(\mathcal{S})\not \subseteq A^c}
  ).$$

As described above, the IM framework shares the use of auxiliary
random variables to represent the sampling model with R.A. Fisher's
fiducial argument. However,
IM emphasizes valid (frequency calibrated) and efficient
probabilistic inference, avoiding the difficulties of R. A. Fisher's
fiducial argument and related theories
\citep{zhang2011dempster,martin2010dempster,liu2015frameworks}.
Together with conditional IMs \citep{martin2015conditional} and
marginal IMs \citep{martin2015marginal}, this theory also provides a
new mathematical tool to develop exact and efficient frequentist methods.
For example, plausibility regions for unknown parameters are exact
Neyman-Pearson confidence regions \citep{martin2013inferential}, and
the plausibility for an assertion of
interest provides
a probabilistic alternative to
Fisher's P-value \citep{martin2014note}. A relevant recent review
of IM methods may be found in \citet{martinPossibilisticInferentialModels2026}.

In this paper, we focus on point estimation. Specifically, the
estimate of $\theta$ is defined as the set of values that attain
maximum plausibility. Additional considerations for selecting a
single representative point, primarily for ease of comparison with
existing methods, are discussed throughout.

\subsection{A predictive random set for ordered uniforms}
\label{s:unif-prs}

In our high-dimensional problems, we require a predictive random set
for sorted uniform random variables to assess parameter plausibility.
Let
\begin{equation}
  \label{eq:s_n}
  \mathbb{S}_n=\{(u_1,\dots,u_n):0<u_1<\cdots<u_n<1\}
\end{equation}
be the sample space of the order statistics
$U=(U_{(1)},\dots,U_{(n)})$ from an i.i.d.\
$\Unif{0,1}$ sample with distribution $\mathbb{U}_n$.
Following the reweighted Anderson--Darling statistic $R_n^2$ in
\citet{liu2023reweighted}, we define the predictive random set
\begin{equation}\label{eq:prs}
  \mathcal{U}
  =
  \{u\in\mathbb{S}_n:\prsbound(u)\leq \prsbound(U)\},
\end{equation}
where $U\sim \mathbb{U}_n$ is an independent draw and
\begin{equation}\label{eq:boundary}
  \prsbound(u)
  =
  R_n^2(u)
  =
  -2C_n\sum_{i=1}^n
  \left[
    \frac{1}{1-\mu_i}\log\frac{u_i}{\mu_i}
    +
    \frac{1}{\mu_i}\log\frac{1-u_i}{1-\mu_i}
  \right],
\end{equation}
with $\mu_i=\frac{i}{n+1}$ and $C_n=\left(2\sum_{k=1}^n\frac{1}{k}\right)^{-1}$.
The statistic is zero when $u_i = \mu_i$ for all $i$ and increases as
the ordered-uniform vector moves
into atypical tail configurations. The normalizing constant $C_n$
matches the convention in \citet{liu2023reweighted}, where it is chosen so
that $E\{R_n^2(U)\}$ is approximately one under the ordered-uniform
law. The predictive random set in \eqref{eq:prs} is therefore the
nested random level set
of the reweighted Anderson--Darling statistic, with smaller boundary
values corresponding to more typical auxiliary configurations.

This construction mirrors the classical one-sample goodness-of-fit
problem. The random threshold
$R^2_n(U)$ ensures the random set is calibrated for the distribution
$\mathbb{U}_n$. While this method checks for matches to an
i.i.d.~uniform sample, the specific sensitivity to distributional
features depends on the chosen association.

\section{Estimators for the classical high-dimensional normal mean
problem}\label{s:cmnm}

\subsection{Generalized probability integral transform for order
statistics from non-i.i.d.\ models}
\label{s:gpit}

In this subsection, we construct an IM for the
high-dimensional normal mean problem using a GPIT
for order statistics along with the predictive random
set construction in
Section~\ref{s:unif-prs}. To capture
global characteristics through the lens of the ordered sample, we
relate the distribution of the order
statistics ${X_{(1)}<\cdots<X_{(n)}}$ to a sample of uniform
order statistics
$U_{(1)}<\cdots<U_{(n)}$ taking values in
$\mathbb{S}_n$. Throughout this section, we write ${(Y_i, Y_{i+1}, \dots, Y_k)}$ as
$Y_{i:k}$ and ${(Y_{(i)},
Y_{(i+1)}, \dots, Y_{(k)})}$ as $Y_{(i:k)}$.
When $X_{1:n}$ share a common continuous CDF $F$,
${\bigl(F(X_{(1)}),\dots,F(X_{(n)})\bigr) \sim \mathbb{U}_n}$.
However, when $X_i \sim F_i$ are independent, non-i.i.d., as in
the high-dimensional normal mean problem, the distribution of each
$X_{(k)}|X_{(1:k-1)}$ depends on all of the constituent
distributions, $(F_1, \dots, F_n)$. We seek a transformation such that
applying it to the data under the true $\theta$ yields random variables
that are distributed equivalently to uniform order statistics. Additionally, 
we want this transformation to preserve the geometry of the sample just as the classical PIT would for an i.i.d.~sample. That is, deviations from the true parameter must translate directly to appropriate, measurable deviations within the uniform simplex.

To achieve this, we develop a Rosenblatt transformation-based
approach that factors the joint density of $X_{(1:n)}$ as a product of a 
marginal and ${n-1}$ conditional densities. Let
$C_{k+1,\theta}(x_{(k+1)}|x_{(1:k)})$
be the conditional CDF of $X_{(k+1)} | X_{(1:k)}$ under the model
$X_i \sim N(\theta_i, 1)$ for $i=1,\dots,n$.
Applying the probability integral transform
to each conditional factor gives ${C_{k+1,\theta}(X_{(k+1)} \mid
X_{(1:k)}) \sim \Unif{0,1}}$. However, simply sorting these uniforms
and inserting them into $B(\cdot)$ may shuffle the value, breaking the geometric integrity of the sample.
To preserve the ordering, and thus the geometric structure, of the sample,
we raise the conditional survival ${1 - C_{k+1,\theta}(X_{(k+1)}
\mid X_{(1:k)})}$ to the power
${1/(n-k)}$, producing a $\operatorname{Beta}(1,n-k)$ variable whose
distribution matches the spacing ${(U_{(k+1)} - U_{(k)})/(1 -
U_{(k)})}$ of uniform order statistics.
Placing this correct spacing relative to the previously
computed transform value builds the output in sorted
order at each step, using only the current
conditional CDF. The formal recursive construction is given in
Theorem~\ref{thm:gpit-distributional} below. In this construction,
the use of the survival function is necessary to correctly align the boundaries
of the conditional domains, ensuring that as $X_{(k+1)}$ approaches its
lower limit of $X_{(k)}$, the transformed variable $U_{(k+1)}$ correctly
approaches its corresponding lower limit of $U_{(k)}$.

The component transforms $F_\theta^{(k)}$ are constructed recursively
in Theorem~\ref{thm:gpit-distributional} below. Sequentially, they
define a function $T_\theta(x_{(1:n)})
  =
  \bigl(F_\theta^{(1)}(x_{(1)}),\dots,F_\theta^{(n)}(x_{(1:n)})\bigr)$
which maps the ordered sample
to the uniform order statistic distribution at the true $\theta$.
Furthermore, Lemma~\ref{lem:gpit-characterization} shows that
$T_\theta$ is a bijection from
$\mathcal{X}_{<}$ onto $\mathbb{S}_n$ and that the distributional
equivalence characterizes
the model. The bijection property ensures that solving $T_\theta(x) = u$ for $x$
is well-posed at every $u \in \mathbb{S}_n$. Corollary~\ref{cor:gpit-reduction} identifies the case in which the
construction reduces to the ordinary probability integral transform. We end with Theorem~\ref{thm:gpit-validity} establishing the validity and efficiency of this procedure.
Proofs of all four results are collected in Appendix~\ref{app:gpit-proofs}.

\begin{theorem}[Distributional equivalence of the generalized transform]
  \label{thm:gpit-distributional}
  Define the transformations recursively. For $k=1,\dots,n-1$, let
  \begin{align*}
    F_\theta^{(k+1)}(x_{(1:k+1)})
    &=
    F_\theta^{(k)}(x_{(1:k)})  \\
    &\quad+
    \{1-F_\theta^{(k)}(x_{(1:k)})\}
    \left[
      1-
      \{1-C_{k+1,\theta}(x_{(k+1)}\mid x_{(1:k)})\}^{1/(n-k)}
    \right] \nonumber
  \end{align*}
  where $F_\theta^{(1)}(x_1)
    =
    1-\{1-F_{X_{(1)}}(x_1;\theta)\}^{1/n}.$
  Then, under the true value of $\theta$, $$\left(
      F_\theta^{(1)}(X_{(1)}),
      \dots,
      F_\theta^{(n)}(X_{(1:n)})
    \right)
    \stackrel{d}{=}
    U_{(1:n)}$$ where $U_{(1:n)}\sim \mathbb{U}_n$.
  
\end{theorem}

The goodness-of-fit based predictive random set can be applied to
$$\{
    F_\theta^{(1)}(X_{(1)}),\dots,
    F_\theta^{(n)}(X_{(1:n)})
  \}$$
as a sorted sample from $\Unif{0,1}$. Evaluating these quantities
requires the sequential conditional CDFs of the order statistics. For
the minimum, $F_{X_{(1)}}(x_{(1)};\theta) = 1 - \prod_{i=1}^n [1 -
\Phi(x_{(1)} - \theta_i)]$. For higher-order statistics, conditioning
on $X_{(1:k-1)} = x_{(1:k-1)}$ determines the values of the smallest
$k-1$ observations but not the assignment of parameters to ordered
positions, so the conditional distribution involves a sum over all
compatible parameter assignments. Explicit assignment-sum formulas
for all sequential conditional CDFs and the corresponding gradients
of the transform components are collected in
Appendix~\ref{app:gpit-computation}.

\begin{lemma}[Bijection characterization of $T_\theta$]
  \label{lem:gpit-characterization}
  The map $T_\theta$ is a one-to-one lower-triangular map from
  $\mathcal X_< = \{x_{(1:n)} : x_{(1)} < \cdots < x_{(n)}\}$ onto $\mathbb{S}_n$.
  Consequently, for any random vector
  $Y = Y_{(1:n)}$ taking values in $\mathcal X_<$, $Y
    \stackrel{d}{=}
    X_{(1:n)}$
  under the model $X_i\sim N(\theta_i,1)$ independently if and only if
    $T_\theta(Y)
    \stackrel{d}{=}
    U_{(1:n)}$.
\end{lemma}

\begin{corollary}[Reduction to the probability integral transform]
  \label{cor:gpit-reduction}
  If $X_1,\dots,X_n$ are i.i.d.\ with common continuous CDF $F$, then for $k=1,\dots,n$, we have that $F_\theta^{(k)}(x_{(1:k)})=F(x_{(k)})$. Thus the construction reduces to the ordinary probability integral transform
  applied to the order statistics.
\end{corollary}

Corollary~\ref{cor:gpit-reduction} provides an analytically checkable
identity that can be used to validate numerical implementations of the GPIT before applying them to the non-i.i.d.\ setting.
The bijection of Lemma~\ref{lem:gpit-characterization} and the distributional
result of Theorem~\ref{thm:gpit-distributional} together guarantee that the
IM induced by the GPIT is both valid and efficient in the sense of \eqref{eq:valid-prs}. For observed data $X$, we write the plausibility of a candidate $\theta$ as
\begin{equation}\label{eq:gpit-pl}
  \operatorname{pl}_X(\theta)=P\bigl\{B(U)\geq B(T_\theta(X))\bigr\},
\end{equation}
where $U\sim\mathbb{U}_n$ is an independent draw and $B=R_n^2$ is the statistic
\eqref{eq:boundary}.

\begin{theorem}[Validity and efficiency of the GPIT inferential model]
  \label{thm:gpit-validity}
  Under the true value of $\theta$, the predictive random set $\mathcal{U}$ in \eqref{eq:prs} is efficient
  for the GPIT auxiliary variable in that $\operatorname{Pr}\bigl[\operatorname{pl}_X(\theta)\ge
    1-\alpha\bigr]=\alpha$ for all $\alpha\in(0,1)$
\end{theorem}

\subsection{Point estimation}
\label{s:point-estimation}

To achieve point estimates, we proceed in two steps. First, we
determine a maximally plausible set-value for $\theta$ using the
association in Section~\ref{s:gpit}. Then, we associate each
$\theta_i$ to a corresponding $X_j$ using the conditional
distribution of $\theta_i$ given the set value $\theta$ and $X$. For
observed data $X$, the plausibility of a candidate $\theta$ is given by
\eqref{eq:gpit-pl}. A small value of
$B(T_\theta(X))$ signals that the transformed data lies near the
center of the ordered-uniform distribution, corresponding to high
plausibility. The maximum plausibility estimator therefore solves
\begin{equation}\label{eq:mpe-obj}
  \hat\theta_{\rm
  MP}=\arg\min_{\theta\in\mathbb{R}^n}\,B\bigl(T_\theta(x_{(1:n)})\bigr).
\end{equation}
The estimator selects the $\theta$ that makes the
GPIT output as
close as possible
to the center of $\mathbb{S}_n$. The assignment-sum formulas that define
$T_\theta(x_{(1:n)})$ sum over all possible assignments of
$\theta_1,\dots,\theta_n$
to the ordered positions, so $T_\theta(x_{(1:n)})$ depends on $\theta$ only
through the unassigned set $\theta_{(1:n)}$. The objective
\eqref{eq:mpe-obj} is therefore invariant to relabeling of the $\theta_i$,
and optimization recovers the sorted values
$\vartheta_k=[\hat\theta_{\rm MP}]_{(k)}$
for $k=1,\dots,n$. Gradient-based minimization is practical because both
$T_\theta(x_{(1:n)})$ and its gradient admit closed-form sequential evaluation.
Writing $T_\theta(x_{(1:n)}) = (G_1(\theta),\dots,G_n(\theta))$, the gradient
of the objective~\eqref{eq:mpe-obj} follows from the chain rule:
\begin{equation*}
  \frac{\partial}{\partial\theta_h}
  B\bigl(T_\theta(x_{(1:n)})\bigr)
  =
  \sum_{r=1}^n
  \frac{\partial B}{\partial u_r}\bigg|_{u=T_\theta(x_{(1:n)})}
  \cdot
  \frac{\partial G_r(\theta)}{\partial\theta_h},
\end{equation*}
where $\partial B/\partial u_r$ is the closed-form derivative of
\eqref{eq:boundary}. The sequential computation of $G_1,\dots,G_n$ and
all partial derivatives $\partial G_r/\partial\theta_h$ is detailed in
Appendix~\ref{app:gpit-computation}.

Now, the sampling-without-replacement nature of this setting
is characterized by a latent permutation $\tau=(\tau_1,\dots,\tau_n)$
of $\{1,\dots,n\}$ on the permutation set $\mathcal{P}_n$ such that
given the estimated support
$\vartheta_1<\cdots<\vartheta_n$, individual component estimates
for $\theta_1,\dots,\theta_n$ may be written as
$\theta_i \sim \vartheta_{\tau_i}$ simultaneously for some estimated $\tau$.
Similarly, if $\vartheta$ is known, our sampling model may be
written equivalently as $X_i \stackrel{ind.}{\sim} N(\vartheta_{\tau_j}, 1)$
where $\tau \sim \text{Unif}(\mathcal{P}_n)$. This structure induces a conditional distribution for $\tau$ proportional to the likelihood
of the augmented model
\begin{equation}
\label{eq:likelihood-without-replacement}
  L(\tau;\vartheta,X) = \prod_{j=1}^n \phi(X_j - \theta_j) =
  \prod_{j=1}^n \phi(X_j - \vartheta_{\tau_j}).
\end{equation}

\begin{definition}[Posterior distribution of latent assignments]
  \label{prop:posterior-dist-of-theta}
  Suppose $n\geq 2$ and assume that the values of $\theta$,
  $\{\vartheta_k: \vartheta_k=\theta_{(k)}, k=1,\dots,n\}$, are known.
  Let $\tau=(\tau_1,\dots,\tau_n) \in \mathcal P_n$ and suppose that,
  conditional on $\tau$,
  $X_j\stackrel{\text{ind.}}{\sim} N(\vartheta_{\tau_j},1)$,
  $j=1,\dots,n$. We define the posterior distribution of
  $\tau | X,\vartheta$ as
  \begin{equation}\label{eq:full-cond}
    \mbox{Pr}(\theta_1 = \vartheta_{\tau_1},\dots,\theta_n =
    \vartheta_{\tau_n} \mid X_1,\dots,X_n)
    =\frac{\prod_{j=1}^n\phi(X_j-\vartheta_{\tau_j})}
    {\sum_{\tau'} \prod_{j=1}^n\phi(X_j-\vartheta_{\tau'_j})} \propto
    L(\tau;\vartheta,X),
  \end{equation}
  where the summation is over all permutations $\tau'$.
\end{definition}

A computationally simpler coordinatewise approximation, based only on the
local likelihood contribution from $X_i$, is
\begin{equation}\label{eq:partial-cond}
  \mbox{Pr}(\theta_i = \vartheta_j \mid X_i)
  =\frac{\phi(X_i-\vartheta_j)}
  {\sum_{k=1}^n\phi(X_i-\vartheta_k)}.
\end{equation}
While the full working posterior
\eqref{eq:full-cond} captures the complete dependence structure, its
evaluation remains computationally infeasible for large $n$. For this
reason, we identify
a simpler estimator based on the coordinatewise approximation
\eqref{eq:partial-cond} to be
\begin{equation}\label{eq:mpe-classic-01}
  \hat{\theta}_i
  =\frac{\sum_{k=1}^n \vartheta_k\,\phi(X_i-\vartheta_k)}
  {\sum_{k=1}^n \phi(X_i-\vartheta_k)},
  \qquad i=1,\dots,n.
\end{equation}
Discrepancies between the coordinatewise approximation and the
permuted assignment are likely to be small. Significant error would
only arise in a pathological case where the optimization in
\eqref{eq:mpe-obj} clusters
the $\theta_i$ in a different pattern than the $X_i$ are observed.
Because the optimal $\{\vartheta_k\}$ is determined by the joint distribution
of the order statistics, it is unlikely that the maximization process
would yield a $\theta$ that is significantly misaligned with the observed
clustering of $X$.

Here, $\{\vartheta_k\}$ is obtained from the IM-based
set estimate
constructed earlier, and \eqref{eq:mpe-classic-01} is the corresponding
posterior mean under the coordinatewise working approximation. This estimator
can be interpreted as a practical point summary of the
maximum-plausibility output. We recognize that this approximated
point estimate may not be a solution to
$\eqref{eq:mpe-obj}$. After producing this point estimate, one may
verify the result is reasonable by calculating
$\text{pl}_X(\theta)$ as in \eqref{eq:gpit-pl}. The large-$n$
i.i.d.\ surrogate described next provides a scalable route for obtaining
such summaries when the exact GPIT construction is too
costly.

\subsection{An i.i.d.\ model approximation for large $n$}

The exact GPIT construction for
the classical high-dimensional normal mean problem retains the
finite-population assignment structure of the ordered observations.
The resulting sampling-without-replacement dependence is the source of
the assignment sums in the exact sequential formulas, and it can be
computationally infeasible for large $n$. A scalable surrogate replaces the permutation in \eqref{eq:likelihood-without-replacement} by independent draws from
the same finite set of labels.
\begin{equation}\label{eq:sample-with-replace}
  \tilde X_i=\theta_{I_i}+Z_i, \qquad
  I_i \stackrel{\text{i.i.d.}}{\sim}\Unif{\{1,\dots,n\}}, \qquad
  Z_i \stackrel{\text{i.i.d.}}{\sim}N(0,1).
\end{equation}
Conditional on $\theta$, the $\tilde X_i$'s are i.i.d.\ with CDF
\begin{equation}\label{eq:sample-with-replace-final}
  \tilde F_\theta(x)
  =
  \frac{1}{n}\sum_{j=1}^n \Phi(x-\theta_j).
\end{equation}
Thus the usual ordered-uniform association
$\tilde F_\theta(\tilde X_{(k)})=U_{(k)}$, $k=1,\dots,n$, can be
used with the predictive random set for uniform order statistics
described earlier.
This approximation preserves the marginal distribution of each~$X_i$,
but not the joint distribution. Under the standard model
without replacement,
all labels are distinct with probability~1. Under the surrogate
\eqref{eq:sample-with-replace}
the probability of selecting all~$n$ distinct labels is
$n!/n^n$. For~$m$
sampled labels, the gap decays as $O(m^2/n)$, which justifies the
approximation
when the effective parameter dimension is small. This justification
follows from
finite-exchangeability and finite-population comparisons
\citep{freedman1977remark,diaconis_finite_1980,darroch1988sampling}.
The surrogate is best suited for goodness-of-fit and uniform-spacing criteria
driven by the empirical distribution~$\tilde F_\theta$, not for
label-sensitive inference problems.

\section{Composite predictive random sets}
\label{s:composite}

\subsection{Combining multiple plausibility contours}
\label{s:multiple-coherence}

In some applications, more than one IM association is available for the same
model. When these associations capture
different features of the sample, it is natural to combine their plausibility
contours when forming a maximum plausibility estimate. To cover both a single
association with several derived features and several separate associations,
let $T_j(X)=a_j(\theta,U_j)$ for $j=1,\dots,J$ where $U_j\sim P_{U_j}$, $T_j(X)$ is the feature of the data used in the $j$th plausibility
assessment, $a_j$ is the corresponding association, and $P_{U_j}$ is the
distribution of the auxiliary variable $U_j$ on $\mathcal U_j$. Let
$\mathcal S_j$ be a valid predictive random set for $U_j$. For each candidate
$\theta$, define the pointwise plausibility $q_j(\theta;X) \equiv \operatorname{pl}_{j,X}(\{\theta\})$
where $\operatorname{pl}_{j,X}(A)$ is the plausibility of
$A \subset \mathcal{U}_j$ under association $a_j$.
Large values of $q_j(\theta;X)$ indicate that the $j$th association
explains the observed data under the fitted value $\theta$ without forcing
all compatible auxiliary values into an atypical region of their distribution.

To combine the plausibilities from $J$ associations, we use the
\textit{bottleneck score}, defined as $C(\theta;X) \equiv
  \min_{1\le j\le J} q_j(\theta;X)$. Maximizing $C(\theta;X)$ selects a fitted value whose least favorable
component plausibility is as large as possible. This maximin criterion is
natural when each component plausibility must be acceptably large. The raw
bottleneck score
is an optimization device. It is not, by itself, a calibrated plausibility
contour because the component scores are generally dependent through the
common observed data and the common fitted value. For calibration, let $X'$ be an independent draw from the sampling model
$P_\theta$ at a fixed $\theta\in\Theta$. The calibrated contour induced by
the bottleneck score is $q_{C,X}(\theta)
  =
  P_\theta\{C(\theta;X')\le C(\theta;X)\}$.
This definition uses the joint distribution of the full vector
$(q_1(\theta;X),\dots,q_J(\theta;X))$. In the following proposition (with proof in Appendix \ref{app:valid-combined}), we show that $q_{C,X}(\theta)$ is stochastically no smaller than uniform, and thus supports valid inference.

\begin{proposition}[Validity of inference using $q_{C,X}(\theta)$]
\label{prop:joint-coherence-calibration}
For any fixed $\theta\in\Theta$, the composite contour
$q_{C,X}(\theta)$ is valid, satisfying $
    P_\theta\{
        q_{C,X}(\theta)\le \alpha
    \}
    \le \alpha$ for $0<\alpha<1$.
\end{proposition}

\subsection{Two radial associations for the high-dimensional normal mean}
\label{s:radial-scores}
For the classical high-dimensional normal mean problem, when
utilizing multiple associations,
we use the convention that
the first association corresponds to the ordered-uniform shape
contour, denoted by $q_{\rm sh}(\theta;X)$. Depending on the method, this
is either the exact generalized transform contour in
\eqref{eq:gpit-pl} or its surrogate
counterpart based on the i.i.d.\ transform
\eqref{eq:sample-with-replace-final}. We take the second score to be
one of two radial
contours. The central residual-radius association is induced by the
residual vector $z_X(\theta)=X-\theta$
and its squared radius $R_X^c(\theta)=\|z_X(\theta)\|^2$.
At the true value, $R_X^c(\theta)\sim\chi^2_n$. Let $f_n$ denote the
$\chi^2_n$ density and let $R\sim\chi^2_n$. The corresponding central
radial pointwise plausibility function is $q_{R^c}(\theta;X)
  =
  P\{f_n(R)\le f_n(R_X^c(\theta))\}$.
  
\begin{remark}
  For the maximum likelihood estimator $\hat\theta_{\rm MLE}=X$,
  $R_X^c(\hat\theta_{\rm MLE})=0$. Therefore, $q_{R^c}(\hat\theta_{\rm MLE};X)
    =1$ for $n=1,2$ and $q_{R^c}(\hat\theta_{\rm MLE};X)=0$ for $n\geq 3$. This provides a new structural explanation for Stein's paradox through the lens of auxiliary plausibility.
\end{remark}

The noncentral observed-radius association is based on the statistic
$S_X=\|X\|^2$
and the noncentrality parameter $\lambda(\theta)=\|\theta\|^2$.
If $F_{n,\lambda}$ denotes the distribution function of a noncentral
$\chi^2$ law with $n$ degrees of freedom and noncentrality parameter
$\lambda$, then the observed-radius PIT auxiliary variable is
${U_{R^{nc},X}(\theta)=F_{n,\lambda(\theta)}(S_X)}$.
At the true value, $U_{R^{nc},X}(\theta)\sim\Unif{0,1}$. Using the centered
two-sided predictive random set on $(0,1)$ yields the noncentral radial
pointwise plausibility $\operatorname{q}_{R^{nc}}(\theta;X)
  =
  2\min\{U_{R^{nc},X}(\theta),1-U_{R^{nc},X}(\theta)\}$. The two bottleneck scores used in the numerical comparisons are therefore
\begin{align*}
  C_{R^c}(\theta;X)
  &=
  \min\{q_{\rm sh}(\theta;X),q_{R^c}(\theta;X)\}\\
  C_{R^{nc}}(\theta;X)
  &=
  \min\{q_{\rm sh}(\theta;X),q_{R^{nc}}(\theta;X)\}.
\end{align*}

\section{Estimators for over-parameterized $g$-modeling}\label{s:eb}

In this section, we consider an empirical Bayes sister-problem
to the classical high-dimensional normal mean problem called
$g$-modeling, formulated as the over-parameterized hierarchical
model in \eqref{eq:g-modeling}. We pursue this problem for two
complementary reasons. First, when $n$ is large, the exact IM
optimization in Section~\ref{s:cmnm} involves a prohibitive number of
permutations in assigning parameters to order positions, whereas the
$g$-modeling approach provides a computational surrogate through its
i.i.d.~structure that avoids these combinatorial challenges. Second,
IM-based inference for over-parameterized, hierarchical $g$-modeling itself represents a
novel contribution. This common modeling scenario has not been
addressed within the IM framework, providing an important
methodological advance for general deconvolution problems. We establish an
association based on order statistics and sorted uniforms, define a
valid predictive random set, and develop maximum plausibility
estimation methods that yield both an estimate of $G(\cdot)$ and
subsequent point estimates for $\theta$.

\subsection{Compound decision and empirical-Bayes shape estimation}

The compound-decision framework evaluates estimation of $\theta$ by
average loss across components, indexed by the empirical distribution
$G_{n,\theta}(t)
  =
  \frac{1}{n}\sum_{i=1}^n {\bf 1}\{\theta_i\le t\}$. If $G_{n,\theta}$ is
known, the oracle rule \citep{zhang1997empirical}
$
  t^*_{G_{n,\theta}}(x)
  =
  \frac{\int u\,\phi(x-u)\,dG_{n,\theta}(u)}
  {\int \phi(x-u)\,dG_{n,\theta}(u)}
$
minimizes average squared error loss by coupling component estimates
through their shared distribution. This is analogous to how the order
statistics in Section \ref{s:gpit} incorporate global structure. The
ordinary MLE $\hat\theta = X$ ignores this shape.

The empirical-Bayes and compound decision literature can be read as
attempts to estimate $G_{n,\theta}$ and approximate the oracle rule
(see \citet{zhang2003compound} for a review). A main conceptual lesson
is that the empirical
distribution of the unknown effects provides a benchmark richer than the first
two moments used by linear empirical-Bayes or James--Stein rules.
\citet{efron_two_2014} organizes the main approaches to $g$-modeling,
which estimates $G$ on the $\theta$-scale, and $f$-modeling, which
works on the $x$-scale. On the $g$-modeling side,
\citet{zhang1997empirical} establishes asymptotic minimax
empirical-Bayes estimators for normal means, and
\citet{jiang2009gmleb} advance this via the non-parametric MLE-based generalized
maximum likelihood empirical-Bayes method, achieving
near-oracle risk. \citet{brown2009nonparametric} pursue $f$-modeling
through Tweedie's formula applied to a kernel density estimate of the
marginal, also establishing asymptotic optimality.
\citet{ritov2024oracle} gives a direct game-theoretic proof via the
nonparametric maximum likelihood estimator and Stein's unbiased risk
estimate connection (where the nonparametric maximum likelihood
  estimator is asymptotically minimax in both empirical Bayes and
compound decision problems) without relying on the standard oracle argument. In recent developments, \citet{jiang2025estimation} show that their
auto-modeling approach outperforms
$g$-modeling in terms of squared-error loss. This raises a
natural question for the present work: {\it when $g$-modeling is viewed as a
  surrogate i.i.d.\ model for the unordered component shape, can an
  over-parameterized IM construction improve on the
  nonparametric maximum likelihood estimator by replacing likelihood
maximization with maximum plausibility estimation?}

\subsection{The sampling model and IM association}
\label{s:g-model-association}

Because $G(\cdot)$ is a CDF, and thus a bounded, monotone function,
it can be approximated to arbitrary precision by a step function.
With this justification, and by following the lead of previous work
for $g$-modeling \citep{efron_two_2014,efron2016empirical}, we take a
numerical approach
to approximate $G(\cdot)$ by discretization. Accordingly, with
minimal loss of generality, we assume that $\theta$ is discrete with
known support. Denote by $\vartheta_1 < ... < \vartheta_K$ the known space
of $\theta$ with a large $K>n$, and denote by $\gamma_k$ the unknown
probability mass at
$\vartheta_k$, where
\begin{equation}\label{eq:numeric-G}
  \gamma \equiv (\gamma_1,...,\gamma_K)
  \in \left\{(\gamma_1,...,\gamma_K):
    \gamma_k \geq 0 \mbox{ for all $k=1,...,K$;
  $\sum_{k=1}^K \gamma_k = 1$}\right\}.
\end{equation}
In this scenario, $G(\cdot)$ is represented by the $K$-dimensional
vector $\gamma$ in \eqref{eq:numeric-G}. In terms of
$G(\theta) =
  \sum_{\vartheta_k \le \theta}\gamma_k$,
$X_1,...,X_n$ is a sample of the convolution of
$G(\cdot)$ and $N(0,1)$:
\begin{equation}\label{eq:numeric-C}
  F_{\vartheta, \gamma}(x) = \sum_{k=1}^K \gamma_k \Phi(x-\vartheta_k)
\end{equation}
where $\Phi(\cdot)$ denotes the CDF of the standard normal
distribution $N(0,1)$.
Furthermore, the
$\theta_i$ are conditionally independent of each other given $\gamma$
and $X$ with
\begin{equation}\label{eq:numeric-T}
  \mbox{Pr}(\theta_i = \vartheta_j | \gamma, X)
  = \frac{\gamma_j\phi(X_i-\theta_j)}
  {\sum_{k=1}^K\gamma_k\phi(X_i-\vartheta_k)},
  \qquad j=1,...,K
\end{equation}
where $\phi(\cdot)$ is the probability density function of the
standard normal distribution.

From \eqref{eq:numeric-C} and \eqref{eq:numeric-T}, we
see that inference about $\theta$ is reduced to inference
about $\gamma$.
Additionally, it is known that $F_{\vartheta, \gamma}(X_{i})$'s form a sample
of size $n$ from $U(0,1)$.
For our inferential approach, we take a sorted uniform sample of size $n$,
denoted by $U_{(1)} < \cdots < U_{(n)}$, as the auxiliary variables
and specify the association as
\begin{equation}\label{eq:numeric-association}
  U_{(i)} = F_{\vartheta, \gamma}(X_{(i)})
  = \sum_{k=1}^K \gamma_k \Phi(X_{(i)}-\vartheta_k).
\end{equation}

\subsection{Induced random sets in the $\gamma$ space}

A realization of the random set \eqref{eq:prs} induces the
random sets in the $\gamma$ space
through the convolution \eqref{eq:numeric-association}. With
$m_\gamma(x)
=
\sum_{k=1}^K \gamma_k \Phi(x-\vartheta_k)$,
we have the predictive random set for $\gamma$ as 
$\Gamma(U) =
  \left\{(\gamma_1, ..., \gamma_K):\;
    \prsbound\!\left(
        m_\gamma(X_{(1)}) \\
        \cdots \\
        m_\gamma(X_{(n)})
\right)
    \leq \prsbound(U)
  \right\},$
which defines the set of
numerical approximations to the unknown distribution $G(\cdot)$.
Standard IM uncertainty quantification about
$G(\cdot)$ can be carried out as prescribed in Section~\ref{s:IMs}.
In the context of multiple testing, for example,
inference can be made about the fractional number of outliers.
For the high-dimensional normal mean problem, especially in the
context of Stein's paradox,
inference about $\theta_i$'s is often of interest
and is obtained by using \eqref{eq:numeric-association} and
\eqref{eq:numeric-T}.
This is discussed in detail in the following subsection. 

\subsection{Maximum plausibility estimation}
\label{s:MPE}

A straightforward application of these IMs is point estimation.
Given an estimate of $G$, the usual Bayes posterior mean,
$\hat{\theta}_i(\gamma) =
  \frac{\sum_{k=1}^K \gamma_k\phi(X_i-\vartheta_k) \vartheta_k}
  {\sum_{k=1}^K \gamma_k\phi(X_i-\vartheta_k)}$,
provides the most efficient estimator \citep{brown1971admissible},
in terms of mean square error $\mbox{MSE}(\hat{\theta}) =
  \frac{1}{n}\sum_{i=1}^n(\hat{\theta}_i-\theta_i)^2$. Here, we consider the maximum plausibility estimate of $\gamma$.
The smallest predictive random set with nested focal elements,
obtained with the method of \citet{leaf2012inference} for validity,
is given by $\mathcal{U}_{\min} = \left\{
    u: \prsbound(u) \leq \prsbound_{\min}
  \right\}$
where $\prsbound_{\min} = \min_\gamma
  \prsbound\{
      m_\gamma(X_{(1)}) \dots
      m_\gamma(X_{(n)}\}$
This induces the maximum plausibility set for $\gamma$:
\begin{equation}\label{eq:mpe-gamma}
  \Gamma_{\min} =
  \left\{(\gamma_1, ..., \gamma_K):\;
    \prsbound\left(
        m_\gamma(X_{(1)}),
        \dots,
        m_\gamma(X_{(n)})\right)
    \leq \prsbound_{\min}
  \right\}.
\end{equation}
Therefore, the maximum plausibility
is necessarily one and is achieved for all $\gamma\in \Gamma_{\min}$.
Since $\Gamma_{\min}$ given in \eqref{eq:mpe-gamma} is a set,
we have a set of MPEs of $\theta_i$, one for each $\gamma\in \Gamma_{\min}$.

For simple practical applications, especially when comparing with
other methods, we can summarize the set of MPEs by
two extremes, the lower and upper posterior expectations $\hat{\theta}_i^{(\mbox{lower})} = \min_{
    \gamma \in \Gamma_{\min}
  } \hat{\theta}_i(\gamma)$ and $\hat{\theta}_i^{(\mbox{upper})} = \max_{
    \gamma \in \Gamma_{\min}
  } \hat{\theta}_i(\gamma)$
for ${i=1,...,n}$.
When a single value estimate is desired,
the mean of the two extremes may be used, that is, $\hat{\theta}_i = \frac{
    \hat{\theta}_i^{(\mbox{lower})} +
  \hat{\theta}_i^{(\mbox{upper})}}{2}$.
This approach is used in Section~\ref{s:numerical} for comparing with other methods.
It is worth noting that because this midpoint estimator $\hat\theta_i$
is formed componentwise, the vector
$(\hat\theta_1,\dots,\hat\theta_n)$ need not correspond to any single
$\gamma\in\Gamma_{\min}$. Similarly to estimation in the classical
problem, after computing the midpoint estimates, one may assess their
plausibility by evaluating $\operatorname{pl}_X(\theta)$ as in
\eqref{eq:gpit-pl}. Additionally, the ordered-uniform contour that
defines the maximum plausibility set
$\Gamma_{\min}$ can be paired with an additional auxiliary diagnostic on the
induced point estimate $(\hat\theta_1(\gamma),\dots,\hat\theta_n(\gamma))$.
In that case, the bottleneck score from
Section~\ref{s:multiple-coherence} applies here without modification. The
numerical studies use this same construction when the $g$-modeling IM
estimator is evaluated jointly with an additional plausibility contour.


\section{Numerical study}
\label{s:numerical}

This section presents a numerical study of the proposed
IM methodology for both the classical high-dimensional
normal mean problem and
the over-parameterized empirical Bayes variant. First,
we examine finite-sample performance under representative configurations of
$\theta$ and evaluate the i.i.d.-surrogate approximation
against the exact GPIT.
Second, we compare our proposed estimators with classical and modern competitors.
Performance is measured throughout by empirical MSE, averaged over $M = 500$
independent replicates, with Monte Carlo standard errors (SE) reported in
parentheses.

\subsection{Simulation scenarios}
\label{s:sim-scenarios}

We adopt the three benchmark scenarios of \citet{jiang2025estimation},
designed to capture qualitatively different structural features of $\theta$:
\begin{enumerate} 
\item {\it Single mode}: $\theta_i \stackrel{\text{i.i.d.}}{\sim} N(0, 0.01)$

\item {\it Two-mode:}
      $\theta_i \stackrel{\text{i.i.d.}}{\sim}
      \tfrac{1}{2}N(-2, 0.01) + \tfrac{1}{2}N(2, 0.01)$;

\item {\it Spike-and-slab:}
      ${\theta_i \stackrel{\text{i.i.d.}}{\sim}
      0.9\,\delta_0 + 0.1\,N(-3, 1)}$,
      where $\delta_0$ is a point mass at zero.
      
\end{enumerate}
      In scenario 1, parameters concentrate tightly near zero, which advantages
      global shrinkage. For scenario 2, global
      shrinkage is suboptimal due to the bimodal structure. Scenario 3 combines sparsity with a heavy left tail. The sampling model is $X_i = \theta_i + Z_i$, $Z_i
\stackrel{\text{i.i.d.}}{\sim} N(0,1)$.

\subsection{Methods compared}
\label{s:methods-compared}

The large-$n$ study of Table~\ref{tb:2} compares the MLE, empirical-Bayes,
and surrogate-based procedures, while the small-$n$ study of
Table~\ref{tb:1} additionally includes exact-transform analogues.
Both tables
distinguish a central
residual-radius score, denoted ``$+R^c$'', and a noncentral observed-radius
score, denoted ``$+R^{nc}$''.

The \textit{MLE} is $\hat\theta_{\rm MLE} = X$, the baseline.
The \textit{James--Stein positive-part (JS$^+$)} applies positive-part shrinkage
toward the grand mean $\bar X$ as $\hat\theta_{\rm JS}^+
  = \bar X \mathbf{1}
  + \!\left(1 - \frac{n-3}{\|X - \bar X \mathbf{1}\|^2}\right)_{\!\!+}
  (X - \bar X \mathbf{1})$.
\textit{Exact IM} (Table~\ref{tb:1} only) minimizes
$B\!\left(T_\theta(x_{(1:n)})\right)$
where $T_\theta$ is the exact generalized transform of
Section~\ref{s:gpit}, feasible for $n \in \{3,5,8\}$.
\textit{Exact+$R^c$} augments Exact IM with the central residual-radius score and \textit{Exact+$R^{nc}$} uses the observed-radius PIT contour:
\begin{align*}
  \operatorname{pl}_{R^c}(\theta;X)&=P\{f_n(R) \le f_n(\|X-\theta\|^2)\}\\
  \operatorname{pl}_{R^{nc}}(\theta;X)&=2\min\{F_{n,\lambda(\theta)}(S_X),
  1-F_{n,\lambda(\theta)}(S_X)\},
\end{align*}
where $R\sim\chi^2_n$, $S_X=\|X\|^2$, and $\lambda(\theta)=\|\theta\|^2$. Both use the exact
generalized transform shape score and are reported only in Table~\ref{tb:1}.
\textit{IMs} replaces the exact generalized transform by the i.i.d. surrogate
$\tilde F(x;\theta) = n^{-1}\!\sum_{j=1}^n \Phi(x - \theta_j)$ and minimizes
$B\!\left(\tilde F(x_{(1)};\theta),\ldots,\tilde F(x_{(n)};\theta)\right)$
over $\theta$ with analytic gradients, which
scales to arbitrary $n$ and appears as ``Approx'' in Table~\ref{tb:1}.
\textit{IMs(+$R^c$)} extends IMs with the central $\chi^2_n$
residual-radius
criterion (Section~\ref{s:radial-scores}).  The shape plausibility
$\operatorname{pl}_{\rm sh}(\theta;X) = P\{B(U) \ge B(\tilde u(\theta))\}$
and the radial plausibility
$\operatorname{pl}_{R^c}(\theta;X) = P\{f_n(R) \le f_n(\|X-\theta\|^2)\}$
($R \sim \chi^2_n$, $f_n$ its density) are combined via the maximin score
$$\hat\theta_{\rm IM+R^c}
  \in \argmax_{\theta}\,
  \min\!\left\{
    \operatorname{pl}_{\rm sh}(\theta;X),\;
    \operatorname{pl}_{R^c}(\theta;X)
  \right\}.$$

The alternative \textit{IMs(+$R^{nc}$)} uses the same shape score
together with
the noncentral observed-radius contour
$\hat\theta_{\rm IM+R^{nc}}
  \in \argmax_{\theta}\,
  \min\!\left\{
    \operatorname{pl}_{\rm sh}(\theta;X),\;
    \operatorname{pl}_{R^{nc}}(\theta;X)
  \right\}$,
where
$\operatorname{pl}_{R^{nc}}(\theta;X)=2\min\{F_{n,\lambda(\theta)}(S_X),
1-F_{n,\lambda(\theta)}(S_X)\}$,
$S_X=\|X\|^2$, and $\lambda(\theta)=\|\theta\|^2$. In
Table~\ref{tb:1}, these surrogate refinements are reported as
``Approx+$R^c$'' and ``Approx+$R^{nc}$''.
\textit{IMs(EB)} maximizes plausibility over the simplex
$\Delta_K$ of discrete
mixture weights on a grid of $K=1000$ support points over
$[X_{(1)} + \Phi^{-1}(10^{-4}),\; X_{(n)} + \Phi^{-1}(0.9999)]$,
following Section~\ref{s:eb}.
\textit{$g$-modeling} \citep{efron_two_2014} estimates the marginal density
$f_G(x) = \int \phi(x-t)\,dG(t)$ by penalized MLE via the \texttt{deconvolveR}
package, returning the posterior mean via Tweedie's formula.
\textit{Auto-modeling} \citep{jiang2025estimation} uses the paper's
bootstrap-imputation procedure with $l=n$ support points, $B=K=5$,
and KS-guided selection of the $m$-out-of-$n$ resampling ratio,
returning the posterior mean under the fitted mixture prior.
All methods observe the same $({\mu}, X)$ draws per replicate so that
comparisons are paired.
The prevalence of this problem in the literature provides a substantial number
of potential estimation procedures to compare. Additional comparisons
for the same three generating processes for $\theta$ may be found in
\citet{jiang2025estimation}.

\subsection{Results}
\label{s:results}

\newcommand{\revisitingSmallNTableBody}{%
  {\scriptsize
    \setlength{\tabcolsep}{4pt}%
    \renewcommand{\arraystretch}{0.95}%
    \resizebox{0.98\textwidth}{!}{%
      \begin{tabular}{@{}lccccccccc@{}}
        \hline\hline
        \multirow{2}{*}{\textit{Method}}
        & \multicolumn{3}{c}{$N(0,0.01)$}
        & \multicolumn{3}{c}{$\tfrac{1}{2}N(-2,0.01)+\tfrac{1}{2}N(2,0.01)$}
        & \multicolumn{3}{c}{$0.9\delta_0+0.1N(-3,1)$} \\
        \cline{2-4}\cline{5-7}\cline{8-10}
        & $n=3$ & $n=5$ & $n=8$
        & $n=3$ & $n=5$ & $n=8$
        & $n=3$ & $n=5$ & $n=8$ \\
        \hline
        Exact
        & 0.709 & 0.521 & 0.362
        & 0.969 & 0.878 & 0.762
        & 0.876 & 0.655 & 0.524 \\
        & (0.037) & (0.027) & (0.016)
        & (0.042) & (0.033) & (0.026)
        & (0.044) & (0.029) & (0.022) \\
        Exact+$R^c$
        & 0.773 & 0.476 & 0.291
        & 1.085 & 0.991 & 0.816
        & 0.959 & 0.659 & 0.496 \\
        & (0.045) & (0.030) & (0.012)
        & (0.051) & (0.037) & (0.025)
        & (0.051) & (0.032) & (0.019) \\
        Exact+$R^{nc}$
        & 0.432 & 0.292 & 0.177
        & 0.962 & 0.846 & 0.746
        & 0.657 & 0.495 & 0.422 \\
        & (0.031) & (0.021) & (0.011)
        & (0.041) & (0.031) & (0.024)
        & (0.040) & (0.026) & (0.020) \\
        Approx [IMs]
        & 0.960 & 0.690 & 0.448
        & 1.231 & 1.058 & 0.866
        & 1.143 & 0.855 & 0.661 \\
        & (0.051) & (0.037) & (0.020)
        & (0.052) & (0.040) & (0.032)
        & (0.058) & (0.039) & (0.028) \\
        Approx+$R^c$
        & 0.916 & 0.639 & 0.316
        & 1.122 & 1.023 & 0.837
        & 1.063 & 0.784 & 0.540 \\
        & (0.050) & (0.036) & (0.015)
        & (0.052) & (0.037) & (0.026)
        & (0.055) & (0.035) & (0.022) \\
        Approx+$R^{nc}$
        & 0.460 & 0.308 & 0.182
        & 0.983 & 0.841 & 0.752
        & 0.690 & 0.508 & 0.422 \\
        & (0.033) & (0.022) & (0.012)
        & (0.042) & (0.031) & (0.025)
        & (0.042) & (0.027) & (0.020) \\
        \hline
      \end{tabular}%
    }%
  }%
}

\begin{table}
  \caption{\label{tb:1}MSE (SE) for the exact and i.i.d.-surrogate IM
    methods and their radial refinements.
  All results based on $M=500$ Monte Carlo replicates.}
  \centering
  \revisitingSmallNTableBody
\end{table}

Table~\ref{tb:1} shows that the exact methods consistently achieve lower MSE than approximate models at small $n$, though the noncentral radial refinement ($+R^{nc}$) substantially closes the gap across all scenarios. For large $n$ (Table~\ref{tb:2}), estimator performance depends heavily on the underlying signal structure. For the single-mode scenario, which is the most favorable case for global shrinkage, Auto-modeling, IMs($+R^{nc}$) and JS$^+$ excel. With two modes (Scenario 2), global shrinkage is actively harmful. Auto-modeling and IMs(EB) perform best as $n$ increases, while both radial criteria become counterproductive because they fail to capture the bimodal structure. For the spike-and-slab scenario (Scenario 3), the radial criterion assists at small sample sizes (with IMs($+R^{nc}$) dominating at $n=10$ and $n=20$), but IMs(EB) overtakes it by $n=50$ as the radial contour begins to conflict with the structured non-single-mode signal. Overall, IMs(EB) yields the most stable large-$n$ performance, tracking closely with state-of-the-art Auto-modeling. Among classical formulations, $+R^{nc}$ is strictly superior to $+R^c$, though both fail when the data strongly deviates from being single-mode.

\newcommand{\revisitingLargeNTableBody}{%
  {\scriptsize
    \setlength{\tabcolsep}{3.5pt}%
    \renewcommand{\arraystretch}{0.95}%
    \resizebox{0.98\textwidth}{!}{%
      \begin{tabular}{@{}lccccccccc@{}}
        \hline\hline
        \multirow{2}{*}{\textit{Method}}
        & \multicolumn{3}{c}{$N(0,0.01)$}
        & \multicolumn{3}{c}{$\tfrac{1}{2}N(-2,0.01)+\tfrac{1}{2}N(2,0.01)$}
        & \multicolumn{3}{c}{$0.9\delta_0+0.1N(-3,1)$} \\
        \cline{2-4}\cline{5-7}\cline{8-10}
        & $n=10$ & $n=20$ & $n=50$
        & $n=10$ & $n=20$ & $n=50$
        & $n=10$ & $n=20$ & $n=50$ \\
        \hline
        \multicolumn{10}{l}{\underline{\it EB-type methods}} \\
        $g$-modeling
        & 0.388 & 0.416 & 0.173
        & 0.757 & 0.755 & 0.732
        & 0.570 & 0.525 & 0.379 \\
        & (0.012) & (0.010) & (0.004)
        & (0.017) & (0.013) & (0.009)
        & (0.016) & (0.012) & (0.007) \\
        Auto-modeling
        & 0.167 & 0.099 & 0.054
        & 0.703 & 0.559 & 0.394
        & 0.482 & 0.399 & 0.368 \\
        & (0.008) & (0.004) & (0.002)
        & (0.025) & (0.018) & (0.011)
        & (0.020) & (0.012) & (0.007) \\
        IMs(EB)
        & 0.256 & 0.167 & 0.071
        & 0.713 & 0.559 & 0.402
        & 0.576 & 0.450 & 0.334 \\
        & (0.014) & (0.009) & (0.004)
        & (0.028) & (0.019) & (0.011)
        & (0.025) & (0.015) & (0.008) \\
        \multicolumn{10}{l}{\underline{\it Classical methods}} \\
        MLE
        & 1.003 & 1.025 & 0.997
        & 1.008 & 1.027 & 0.999
        & 1.009 & 1.012 & 1.007 \\
        & (0.019) & (0.015) & (0.009)
        & (0.020) & (0.014) & (0.009)
        & (0.020) & (0.014) & (0.009) \\
        James--Stein (JS$^+$)
        & 0.232 & 0.126 & 0.051
        & 0.869 & 0.867 & 0.818
        & 0.518 & 0.495 & 0.494 \\
        & (0.011) & (0.006) & (0.002)
        & (0.017) & (0.012) & (0.008)
        & (0.018) & (0.012) & (0.007) \\
        IMs
        & 0.374 & 0.224 & 0.091
        & 0.787 & 0.615 & 0.426
        & 0.618 & 0.461 & 0.353 \\
        & (0.018) & (0.011) & (0.004)
        & (0.029) & (0.020) & (0.012)
        & (0.025) & (0.015) & (0.009) \\
        IMs(+$R^c$)
        & 0.270 & 0.181 & 0.075
        & 0.742 & 0.660 & 0.559
        & 0.486 & 0.414 & 0.423 \\
        & (0.011) & (0.008) & (0.004)
        & (0.021) & (0.016) & (0.010)
        & (0.018) & (0.011) & (0.007) \\
        IMs(+$R^{nc}$)
        & 0.148 & 0.096 & 0.041
        & 0.700 & 0.646 & 0.550
        & 0.421 & 0.379 & 0.366 \\
        & (0.010) & (0.006) & (0.002)
        & (0.021) & (0.016) & (0.010)
        & (0.018) & (0.012) & (0.007) \\
        \hline
      \end{tabular}%
    }%
  }%
}

\begin{table}
  \caption{\label{tb:2}MSE (SE in parentheses) for the large-$n$
    comparison, $M=500$ replicates.
  Methods are grouped as EB-type (top) and classical problem (bottom).}
  \centering
  \revisitingLargeNTableBody
\end{table}

\section{Discussion}
\label{s:discussion}

This paper resolves a foundational gap in the classical
multivariate normal means problem by providing a prior-free,
structural explanation for Stein’s paradox.
By analyzing the
radial auxiliary structure within the IM framework, we
demonstrate that for $n\geq 3$, the MLE becomes structurally
implausible. Unlike classical shrinkage and empirical Bayes
methods, this approach captures the global coordinate structure
without imposing external geometric or hierarchical constraints.

To establish this result, we constructed a class of valid and efficient point 
estimators driven by a rank-preserving, bijective GPIT. To ensure scalability 
of this method to high-dimensional settings, we introduced an i.i.d.~sampling-
with-replacement surrogate and developed an IM formulation for over-
parameterized $g$-modeling. Beyond its utility as an approximation, the IM 
formulation of $g$-modeling offers a valuable tool for applications where the 
hierarchical model itself holds scientific meaning, such as in cases of common 
instrument measurement error. Furthermore, we introduced a maximin procedure 
to combine multiple distinct plausibility contours, accounting for differing 
structural features of the data. Numerical experiments establish our IM-based 
estimators outperform classical geometric shrinkage and hierarchical methods.

Several directions for future research remain open. First, because evaluations
of the exact generalized transform and its gradients are computationally
expensive, the development of more scalable search algorithms would make the
exact formulation more widely applicable. Second, although we have concentrated
on point estimation, the same auxiliary associations can be used to construct
plausibility regions and confidence procedures. Finally, our framework suggests
a broader way to analyze shrinkage and other regularized estimators. Rather
than evaluating these methods solely through risk, one can investigate the
plausibility of the auxiliary variables they induce under the fitted model to
gain further insight into regularized statistical inference.

\section{Acknowledgments}
This work is supported by funds from the National
Science Foundation \mbox{(NSF Grant DMS-2412629)}.

\bibliography{mnms}

\appendix

  \section{Proofs for Section~\ref{s:gpit}}
  \label{app:gpit-proofs}

  \begin{proof}[Proof of Theorem~\ref{thm:gpit-distributional}]
    Let
    $
      C_1=F_{X_{(1)}}(X_{(1)};\theta).
    $
    Since $X_{(1)}$ has a continuous distribution,
    $
      C_1\sim \Unif{0,1}.
    $
    Therefore
    $
      F_\theta^{(1)}(X_{(1)})
      =
      1-(1-C_1)^{1/n}
    $
    has the $\operatorname{Beta}(1,n)$ distribution, which is the distribution
    of the first order statistic $U_{(1)}$ from an i.i.d.\
    $\Unif{0,1}$ sample. Now suppose that, for some $k\ge1$,
    $
      \left(
        F_\theta^{(1)}(X_{(1)}),
        \dots,
        F_\theta^{(k)}(X_{(1:k)})
      \right)
      \stackrel{d}{=}
      U_{(1:k)}.
    $
    Define the conditional PIT variable
    $
      C_{k+1}
      =
      C_{k+1,\theta}
      (X_{(k+1)}\mid X_{(1:k)}).
    $
    By the conditional probability integral transform,
    $
      C_{k+1}\sim \Unif{0,1}
    $
    and $C_{k+1}$ is independent of
    $\sigma\{X_{(1:k)}\}$. Hence it is independent of
    $
      \left(
        F_\theta^{(1)}(X_{(1)}),
        \dots,
        F_\theta^{(k)}(X_{(1:k)})
      \right).
    $
    Equivalently,
    $
      W_{k+1}=1-C_{k+1}
    $
    is also $\Unif{0,1}$ and independent of the previous
    transformed components. Define
    $
      V_{k+1}
      =
      1-W_{k+1}^{1/(n-k)}.
    $
    Then
    $
      V_{k+1}\sim \operatorname{Beta}(1,n-k),
    $
    and $V_{k+1}$ is independent of the previous transformed components. The recursive definition gives $F_\theta^{(k+1)}(X_{(1:k+1)}) =
      F_\theta^{(k)}(X_{(1:k)}) +
      {\{1-F_\theta^{(k)}(X_{(1:k)})\}V_{k+1}}$.
    Additionally, uniform order statistics satisfy the standard sequential
    representation
    $
      U_{(k+1)}
      =
      U_{(k)}
      +
      (1-U_{(k)})V_{k+1},
    $
    where $V_{k+1}\sim \operatorname{Beta}(1,n-k)$ is independent of
    $U_{(1:k)}$. Therefore
    $
      \left(
        F_\theta^{(1)}(X_{(1)}),
        \dots,
        F_\theta^{(k+1)}(X_{(1:k+1)})
      \right)
      \stackrel{d}{=}
      U_{(1:k+1)}.
    $
    The result follows by induction.
  \end{proof}

  \begin{proof}[Proof of Lemma~\ref{lem:gpit-characterization}]
    The recursive formula shows that
    $T_\theta(x)\in\mathbb{S}_n$ for every $x\in\mathcal X_<$. At each step,
    the conditional CDF value lies in $(0,1)$, so
    $
      F_\theta^{(k+1)}(x_{1:k+1})
      =
      F_\theta^{(k)}(x_{1:k})
      +
      \{1-F_\theta^{(k)}(x_{1:k})\}v_{k+1}
    $
    for some $v_{k+1}\in(0,1)$. Hence the transformed coordinates are strictly
    increasing and remain in $(0,1)$. For $u\in\mathbb{S}_n$, define
    $
      c_1=1-(1-u_1)^n .
    $
    Then the first coordinate of the inverse map is
    $
      x_1=F_{X_{(1)}}^{-1}(c_1;\theta).
    $
    For $k=1,\dots,n-1$, once $x_{1:k}$ have been recovered, set
    $
      v_{k+1}
      =
      \frac{u_{k+1}-u_k}{1-u_k}
      \qquad\text{and}\qquad
      c_{k+1}
      =
      1-(1-v_{k+1})^{n-k}.
    $
    Since $0<u_1<\cdots<u_n<1$, each $v_{k+1}\in(0,1)$. The conditional
    distribution $C_{k+1,\theta}(\cdot\mid x_{1:k})$ is continuous and
    strictly increasing on its support under the normal model, so the next
    coordinate is recovered uniquely as
    $
      x_{k+1}
      =
      C_{k+1,\theta}^{-1}
      (c_{k+1}\mid x_{1:k}).
    $
    Thus $T_\theta$ has a measurable lower-triangular inverse on
    $\mathbb{S}_n$. Let $X^\theta_{(1:n)}$ denote the order statistics generated from
    independent $N(\theta_i,1)$ variables.
    Theorem~\ref{thm:gpit-distributional} gives
    $
      T_\theta(X^\theta_{(1:n)})
      \stackrel{d}{=}
      U_{(1:n)}.
    $
    Since $T_\theta$ is one-to-one, $X^\theta_{(1:n)}$ has the same
    distribution as $T_\theta^{-1}(U_{(1:n)})$. Therefore, for any
    ordered random vector $Y$, if
    $
      T_\theta(Y)\stackrel{d}{=}U_{(1:n)},
    $
    then
    $
      Y
      =
      T_\theta^{-1}\{T_\theta(Y)\}
      \stackrel{d}{=}
      T_\theta^{-1}(U_{(1:n)})
      \stackrel{d}{=}
      X^\theta_{(1:n)}.
    $
    The converse is exactly the distributional statement of
    Theorem~\ref{thm:gpit-distributional}. This establishes the equivalence.
  \end{proof}

  \begin{proof}[Proof of Corollary~\ref{cor:gpit-reduction}]
    Suppose $X_1,\dots,X_n$ are i.i.d.\ with common continuous CDF $F$. Then
    $
      F_{X_{(1)}}(x)=1-\{1-F(x)\}^n,
    $
    so
    $
      F_\theta^{(1)}(x_1)
      =
      1-\{1-F_{X_{(1)}}(x_1)\}^{1/n}
      =
      F(x_1).
    $
    Assume that
    $
      F_\theta^{(k)}(x_{1:k})=F(x_k).
    $
    Given $X_{(k)}=x_k$, the remaining $n-k$ observations exceed $x_k$;
    after truncation, their conditional survival beyond $x_{k+1}>x_k$ is
    $
      \frac{1-F(x_{k+1})}{1-F(x_k)}.
    $
    Hence
    $
      1-
      C_{k+1,\theta}(x_{k+1}\mid x_{1:k})
      =
      \left\{
        \frac{1-F(x_{k+1})}{1-F(x_k)}
      \right\}^{n-k}.
    $
    Substituting this into the recursion gives
      $F_\theta^{(k+1)}(x_{1:k+1})
      =
      F(x_k)
      +
      \{1-F(x_k)\}
      \left[
        1-
        \frac{1-F(x_{k+1})}{1-F(x_k)}
      \right] =
      F(x_{k+1})$.
    Thus, by induction,
    $
      F_\theta^{(k)}(x_{1:k})=F(x_k)$ for $k=1,\dots,n.
    $
    Consequently,
    $
      T_\theta(X_{(1:n)})
      =
      \left(
        F(X_{(1)}),\dots,F(X_{(n)})
      \right),
    $
    which is the classical probability integral transform applied to the order
    statistics.
  \end{proof}

  \begin{proof}[Proof of Theorem~\ref{thm:gpit-validity}]
    By Theorem~\ref{thm:gpit-distributional}, under the true $\theta$ the
    transform output satisfies $T_\theta(X)\stackrel{d}{=}U$, where
    $U\sim\mathbb{U}_n$. Therefore $B(T_\theta(X))\stackrel{d}{=}B(U)$, with
    $U$ and the independent draw $U'$ in \eqref{eq:gpit-pl} both distributed
    as $\mathbb{U}_n$. Since $B$ is continuous on $\mathbb{S}_n$ and $U,U'$
    are independent and identically distributed,
    $P\{B(U')\ge B(U)\}\sim\operatorname{Unif}(0,1)$ by the probability
    integral transform applied to $B(U)$. The calibration identity follows
    immediately.
  \end{proof}

  \section{Sequential GPIT Computation and Gradients}
  \label{app:gpit-computation}

  The goal of this section is to compute $T_\theta(x_{(1:n)}) =
  (G_1(\theta),\dots,G_n(\theta))$ and its gradient for a given $\theta$,
  as required by the minimization~\eqref{eq:mpe-obj}.  The computation
  proceeds by the following recursion in $r = 1,\dots,n$.

  \begin{enumerate}
    \item \textbf{Base case ($r = 1$).}
      The first component is
      $G_1(\theta) = 1 - \{\prod_{i=1}^n \bar F_i(x_1)\}^{1/n}$,
      the probability integral transform of the sample minimum under the
      non-i.i.d.\ model.

    \item \textbf{Recursion ($r = 2,\dots,n$).}
      Set $m = r-1$.  Form two sums: $D_m$, the unnormalized joint
      density of the first $m$ order statistics $X_{(1:m)} = x_{1:m}$, and
      $N_r$, the same sum with the survival threshold raised from $x_m$ to
      $x_r$.  Their ratio $L_r = N_r/D_m$ is the conditional survival
      probability $\Pr_\theta(X_{(r)} > x_r \mid X_{(1:m)} = x_{1:m})$.
      Setting $M_r = L_r^{1/(n-r+1)}$ yields
      $
        G_r = 1-(1-G_{r-1})\,M_r.
      $

    \item \textbf{Gradients.}
      Differentiating the recursion propagates the gradient forward: each step
      requires $\partial D_m/\partial\theta_k$ and
      $\partial N_r/\partial\theta_k$, which are the same sums weighted by
      a per-term log-derivative factor $\ell_{\pi k}^{(m)}$.
  \end{enumerate}

  Theorem~\ref{thm:gpit-conditional-distributions} derives the conditional
  CDF formulas and justifies the sums $D_m$ and $N_r$.
  Theorem~\ref{thm:gpit-gradients} derives the gradient recursion and the
  derivative formulas for $D_m$ and $N_r$.
  These formulas follow from the standard representation of order statistics
  from independent non-i.i.d.~samples; see
  \citet{vaughan1972permanent}, \citet{bapat1989order}, and the survey
  \citet{balakrishnan2007permanents}.
  The i.i.d.\ reduction is the usual conditional distribution of order
  statistics; see \citet{arnold1992first} or \citet{david2003order}. Throughout the section, let $X_1,\dots,X_n$ be independent with
  $X_i\sim N(\theta_i,1)$, and let $x_1<\cdots<x_n$ be fixed ordered
  values. Write
  $
    f_i(t)=\phi(t-\theta_i),
    \qquad
    \bar F_i(t)=1-\Phi(t-\theta_i),
    \qquad
    \rho_i(t)=\frac{f_i(t)}{\bar F_i(t)}.
  $
  Let $\mathcal P_m$, $1\leq m\leq n$, denote the set of ordered
  length $m$-subsets of distinct
  indices from $\{1,\dots,n\}$.
  For $\pi=(\pi_1,\dots,\pi_m)\in\mathcal P_m$,
  write $\{\pi\}^c=\{1,\dots,n\}\setminus\{\pi_1,\dots,\pi_m\}$ and define
  $
    A_\pi^{(m)}(x_1,\dots,x_m)
    =
    \prod_{j\in \pi} f_{\pi_j}(x_j)$ and $
    B_\pi^{(m)}(t)
    =
    \prod_{j\in\{\pi\}^c}\bar F_j(t).
  $
  Here $A_\pi^{(m)}$ is the joint density arising when $X_{\pi_j}=x_j$
  for each $j=1,\dots,m$, and $B_\pi^{(m)}(t)$ is the probability that
  every remaining variable exceeds $t$. For $m=1,\dots,n-1$,
  summing over $\mathcal{P}_m$ gives
  $
    D_m
    =
    \sum_{\pi\in\mathcal P_m}
    A_\pi^{(m)}(x_1,\dots,x_m)B_\pi^{(m)}(x_m),
  $
  the unnormalized joint density of $X_{(1:m)}=x_{1:m}$.
  For $r=2,\dots,n$, let $m=r-1$ and define the corresponding numerator
  $
    N_r
    =
    \sum_{\pi\in\mathcal P_m}
    A_\pi^{(m)}(x_1,\dots,x_m)B_\pi^{(m)}(x_r),
  $
  the same sum, but with the survival threshold raised from $x_m$ to $x_r$.

  \begin{theorem}[Sequential conditional distributions of order statistics]
    \label{thm:gpit-conditional-distributions}
    The following formulas hold. (1) The marginal CDF of the minimum is
        $
          F_{X_{(1)}}(x_1;\theta)
          =
          1-\prod_{i=1}^n \bar F_i(x_1).
        $; (2) For $r=2,\dots,n$, with $m=r-1$,
        \begin{align*}
          &F_{X_{(r)}\mid X_{(1)},\dots,X_{(r-1)}}
          (x_r\mid x_1,\dots,x_m;\theta) \\
          &\qquad
          =
          1-\frac{N_r}{D_m}
          =
          1-
          \sum_{\pi\in\mathcal P_m}
          w_\pi(x_1,\dots,x_m)
          \prod_{j\in\{\pi\}^c}
          \frac{\bar F_j(x_r)}{\bar F_j(x_m)},
        \end{align*}
        where
        $
          w_\pi(x_1,\dots,x_m)
          =
          A_\pi^{(m)}(x_1,\dots,x_m)B_\pi^{(m)}(x_m) / D_m.
        $; and (3) If $\theta_1=\cdots=\theta_n=\theta_0$, then
        $
          F_{X_{(r)}\mid X_{(r-1)}}
          (x_r\mid x_{r-1};\theta_0)
          =
          1-
          \left[
            \frac{1-\Phi(x_r-\theta_0)}
            {1-\Phi(x_{r-1}-\theta_0)}
          \right]^{n-r+1}.
        $
    \begin{proof}
      The result for $X_{(1)}$ follows immediately from independence:
      $
        \Pr(X_{(1)}>x_1)
        =
        \prod_{i=1}^n \bar F_i(x_1).
      $ Now fix $r\ge2$ and write $m=r-1$. Conditioning on
      $
        X_{(1)}=x_1,\dots,X_{(m)}=x_m
      $
      does not identify which original variables produced these ordered values.
      For an assignment $\pi=(\pi_1,\dots,\pi_m)\in\mathcal P_m$, the density
      contribution of the first $m$ ordered observations is
      $
        A_\pi^{(m)}(x_1,\dots,x_m)
        =
        \prod_{q=1}^m f_{\pi_q}(x_q).
      $
      Because $x_m$ is the $m$th order statistic, all unassigned
      variables must
      exceed $x_m$, contributing $B_\pi^{(m)}(x_m)$. Summing over
      all assignments
      gives $D_m$. Similarly, for $x_r\ge x_m$, the event
      $X_{(r)}>x_r$ requires all
      unassigned variables to exceed $x_r$, contributing $B_\pi^{(m)}(x_r)$.
      Summing over assignments gives $N_r$. Hence
      $
        \Pr_\theta\{X_{(r)}>x_r\mid X_{(1)}=x_1,\dots,X_{(m)}=x_m\}
        =
        \frac{N_r}{D_m},
      $
      which proves the conditional distribution formula. Dividing
      each assignment
      contribution by $D_m$ gives the equivalent weighted representation.

      If $\theta_1=\cdots=\theta_n=\theta_0$, all assignment terms
      are identical
      up to the common factor $\prod_{q=1}^m\phi(x_q-\theta_0)$. Therefore
      $
        \frac{N_r}{D_m}
        =
        \left[
          \frac{1-\Phi(x_r-\theta_0)}
          {1-\Phi(x_m-\theta_0)}
        \right]^{n-m}
        =
        \left[
          \frac{1-\Phi(x_r-\theta_0)}
          {1-\Phi(x_{r-1}-\theta_0)}
        \right]^{n-r+1},
      $
      which gives the i.i.d.\ reduction.
    \end{proof}
  \end{theorem}

  \begin{theorem}[GPIT components and gradient formulas]
    \label{thm:gpit-conditional-gradients}
    \label{thm:gpit-gradients}
    Let $G_r(\theta)=F_\theta^{(r)}(x_1,\dots,x_r)$ for $r=1,\dots,n$ denote the
    GPIT components. Setting $L_r=N_r/D_{r-1}$ and
    $M_r=L_r^{1/(n-r+1)}$, the recursion is
    $
      {G_1(\theta)
      =
      1-
      \left\{\prod_{i=1}^n \bar F_i(x_1)\right\}^{1/n}}$
      and $
      G_r(\theta)
      =
      1-
      \{1-G_{r-1}(\theta)\}M_r$ for
$r=2,\dots,n$.
    The gradients satisfy
    $
      \frac{\partial G_1}{\partial\theta_k}
      =
      -\frac{1-G_1}{n}\rho_k(x_1)
    $,
    and for $r=2,\dots,n$,
    $$
      \frac{\partial G_r}{\partial\theta_k}
      =
      M_r\frac{\partial G_{r-1}}{\partial\theta_k}
      -
      \frac{(1-G_{r-1})M_r}{n-r+1}
      \left\{
        \frac{\partial_{\theta_k}N_r}{N_r}
        -
        \frac{\partial_{\theta_k}D_{r-1}}{D_{r-1}}
      \right\}.
    $$
    For $k=1,\dots,n$, define
    $$
      \ell_{\pi k}^{(m)}(t)
      =
      \begin{cases}
        x_{q_\pi(k)}-\theta_k,
        & k\in\{\pi\},\\
        \rho_k(t),
        & k\notin\{\pi\},
      \end{cases}
    $$
    where $q_\pi(k)$ is the position satisfying $\pi_{q_\pi(k)}=k$;
    equivalently,
    $
      \ell_{\pi
      k}^{(m)}(t)=\partial_{\theta_k}\log\{A_\pi^{(m)}B_\pi^{(m)}(t)\}.
    $
    Then
    $$
      \frac{\partial D_m}{\partial\theta_k}
      =
      \sum_{\pi\in\mathcal P_m}
      A_\pi^{(m)}(x_1,\dots,x_m)B_\pi^{(m)}(x_m)
      \ell_{\pi k}^{(m)}(x_m)
    $$
    and
    $$
      \frac{\partial N_r}{\partial\theta_k}
      =
      \sum_{\pi\in\mathcal P_m}
      A_\pi^{(m)}(x_1,\dots,x_m)B_\pi^{(m)}(x_r)
      \ell_{\pi k}^{(m)}(x_r).
    $$

    \begin{proof}
      The formula for $G_1$ follows by substituting the distribution of the
      minimum into the first GPIT transform. For $r\ge2$, the
      conditional survival
      probability is $L_r=N_r/D_{r-1}$, so the GPIT recursion becomes
      $$
        G_r
        =
        1-(1-G_{r-1})L_r^{1/(n-r+1)}
        =
        1-(1-G_{r-1})M_r .
      $$
      Since
      $
        \frac{\partial}{\partial\theta_k}\bar F_k(x_1)=f_k(x_1),
      $
      we have
      $
        \frac{\partial}{\partial\theta_k}
        \log\left\{\prod_{i=1}^n \bar F_i(x_1)\right\}
        =
        \rho_k(x_1),
      $
      which gives
      $
        \frac{\partial G_1}{\partial\theta_k}
        =
        -\frac{1-G_1}{n}\rho_k(x_1).
      $
      For $r\ge2$, differentiating
      $
        G_r=1-(1-G_{r-1})M_r
      $
      gives
      $
        \frac{\partial G_r}{\partial\theta_k}
        =
        M_r\frac{\partial G_{r-1}}{\partial\theta_k}
        -
        (1-G_{r-1})\frac{\partial M_r}{\partial\theta_k}.
      $
      Since
      $
        \frac{\partial M_r}{\partial\theta_k}
        =
        \frac{M_r}{n-r+1}
        \left\{
          \frac{\partial_{\theta_k}N_r}{N_r}
          -
          \frac{\partial_{\theta_k}D_{r-1}}{D_{r-1}}
        \right\},
      $
      the recursive gradient formula follows. It remains to compute $\partial_{\theta_k}D_m$ and
      $\partial_{\theta_k}N_r$. For a fixed assignment $\pi$, if
      $k\in\{\pi\}$, say $\pi_{q_\pi(k)}=k$, then
      $
        \frac{\partial}{\partial\theta_k}A_\pi^{(m)}
        =
        A_\pi^{(m)}\{x_{q_\pi(k)}-\theta_k\},
        \qquad
        \frac{\partial}{\partial\theta_k}B_\pi^{(m)}(t)=0.
      $
      If $k\notin\{\pi\}$, then
      $
        \frac{\partial}{\partial\theta_k}A_\pi^{(m)}=0,
        \qquad
        \frac{\partial}{\partial\theta_k}B_\pi^{(m)}(t)
        =
        B_\pi^{(m)}(t)\rho_k(t).
      $
      Combining these two cases gives
      $
        \frac{\partial}{\partial\theta_k}
        \{A_\pi^{(m)}B_\pi^{(m)}(t)\}
        =
        A_\pi^{(m)}B_\pi^{(m)}(t)\ell_{\pi k}^{(m)}(t).
      $
      Summing over $\pi\in\mathcal P_m$, with $t=x_m$ for $D_m$ and
      $t=x_r$ for $N_r$, yields the stated formulas.
    \end{proof}
  \end{theorem}

\section{Proof of validity for combined plausibility contours}
\label{app:valid-combined}
\begin{proof}[Proof of Proposition~\ref{prop:joint-coherence-calibration}]
If $X\sim P_\theta$, then $C(\theta;X)$ has the same distribution as
$C(\theta;X')$. Therefore $q_{C,X}(\theta)$ is the
distribution transform of the scalar random variable $C(\theta;X)$, evaluated
at its observed value. For any scalar random variable $V$, the
quantity $P\{V'\le V\}$, with $V'$ an independent copy of $V$, is
stochastically no smaller than a $\Unif{0,1}$ random variable. Applying this
with $V=C(\theta;X)$ gives the stated calibration inequality.
\end{proof}

  \section{Competing interests}
  No competing interest is declared.